\documentclass{acm_proc_article-sp}
\usepackage{mathtools}
\usepackage{algorithmic}
\usepackage{algorithm}
\usepackage{rotating}
\usepackage{subfigure}
\usepackage{verbatim}
\usepackage{graphicx}
\usepackage{epsfig}
\usepackage{sidecap}
\usepackage{caption}
\usepackage{color}
\usepackage{amsfonts}
\usepackage{booktabs}

\newtheorem{DE}{Definition}

\begin{document}

\title{Community Detection from Location-Tagged Networks}

\numberofauthors{1} 
\author{
\alignauthor
 Zhi Liu, Yan Huang\\
			\affaddr{Department of Computer Science and Engineering}\\
		  \affaddr{University of North Texas, Denton, Texas}\\
       \email{zhiliu@my.unt.edu, huangyan@unt.edu}          
}

\maketitle

\begin{abstract}

Many real world systems or web services can be represented as a network such as social networks and transportation networks. In the past decade, many algorithms have been developed to detect the communities in a network using connections between nodes. However in many real world networks, the locations of nodes have great influence on the community structure. For example, in a social network, more connections are established between geographically proximate users. The impact of locations on community has not been fully investigated by the research literature. In this paper, we propose a community detection method which takes locations of nodes into consideration. The goal is to detect communities with both geographic proximity and network closeness. We analyze the distribution of the distances between connected and unconnected nodes to measure the influence of location on the network structure on two real location-tagged social networks. We propose a method to determine if a location-based community detection method is suitable for a given network.  We propose a new community detection algorithm that pushes the location information into the community detection. We test our proposed method on both synthetic data and real world network datasets. The results show that the communities detected by our method distribute in a smaller area compared with the traditional methods and have the similar or higher tightness on network connections.


\end{abstract}



\terms{Algorithm}

\section{Introduction}

Many real world systems or web services can be represented as a network such as social networks, transportation networks, the World Wide Web, and biological networks.  Detecting communities from those networks has received considerable attention and is the main focus of many research efforts in the past decade \cite{clauset2004finding,newman2004finding,2010community,2011classification}. Generally, the goal of community detection is to find the subgraphs with tight internal connection based on node connections, labels of nodes, and the weights derived from data or network structure. Nodes in the same community are closer to each other. Therefore, in the real world, a community represents a group of nodes sharing some similar common friends or features.

However, the formation of many real world networks is greatly influenced by the geographic locations of the nodes which has not been fully investigated by the currently literature.  For example, in a social network, people have a high probability to build a connection with his/her colleague or schoolmate because they know each other or in most cases, they became friends because they are geographically close. Furthermore, some network applications, such as FourSquare, are mostly location-based social networks. The geographic location will play even more importance in the social network structure in these platforms. There are preliminary studies on the relationship between social network structure and geographic distance \cite{2010distance} and \cite{onnela2011geographic}. However, those studies do not push location information into community detection.

We observe that the nodes in a tightly connected community tend to be more close to each other in space as well. Location can have different impact on social networks and the impact can be quantified and used in community detection. Introducing  locations of nodes to community detection can improve the performance of detection on real world networks. In this paper, we propose community detection methods that take the locations of the nodes into consideration with the main goal of improving the quality of the detection results in terms of average internal degree, accuracy, and geographic span of detected communities. Our research is based on the following two premises: (1) Location is an important factor and can greatly influence the connection establishment in many location-tagged networks; (2) For many applications, detecting communities with constrained geographic distribution is important. For example, finding local communities will be useful for arranging meetups of communities with similar interests. Knowing geographically constrained communities with potential interests in certain concert or talk shows can help arranging and scheduling the tours.

\begin{figure}[ht!]
        \centering
        \includegraphics[width=0.85\linewidth]{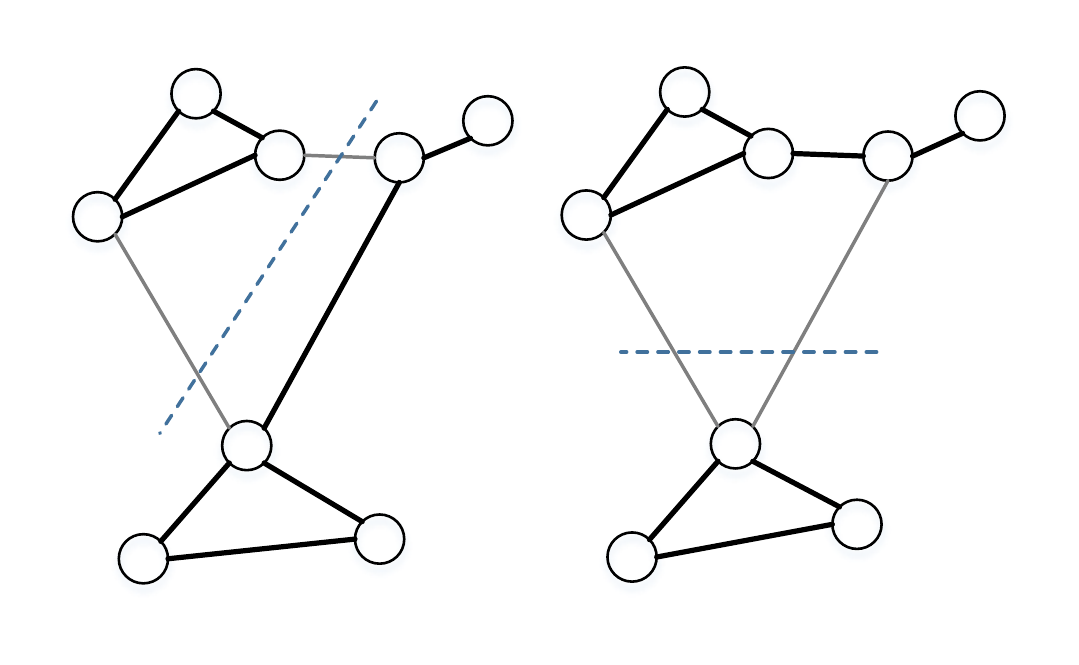}
        \caption{Two different divisions of a small location-tagged network.  The left division is only based on the network structure and the right one takes the locations of the nodes into account.}
        \label{fig:socialreal}
\end{figure}

We focus on finding communities with nodes distributing in a small range of area and at the same time, keeping the connection tightness of the nodes in the community. Figure \ref{fig:socialreal} gives an example of how the geographic location of nodes can influence the detecting results. In this case, we set the number of communities to two. If we only consider the network structure, the left one is a good result. There are only two edges coming across communities. After we introduce the location of the nodes, we will have two communities as in the right side. There are still only two connections across communities however the geographic spans of the two communities are much smaller than the left one. Unfortunately, in some networks, we may need to make a tradeoff on the structure tightness for keeping the nodes in the same community close. This paper presents a way to measure the locality and node similarity and gives an guidance on if a given network has locality in communities.

This paper makes the following contributions:
\begin{itemize}
\item This is the first effort to detect communities with locality on large location-tagged networks;
\item Given a location-tagged network, we proposed a new measurement called Total Variation Difference to help determine if the network has a locality property and a location-based community detection method is suitable. We introduce two concepts: connection locality to measure the closeness of two nodes and node similarity to measure the ``importance'' of an edge. Using these two concepts, we propose a new community detection algorithm that pushes the location information into the community detection; 
\item  We propose optimization techniques and indexing method to allow the algorithm to scale well for large networks. It took around 30 seconds to detect communities from a real network of 20,000 nodes;
\item We test our proposed method on both synthetic data and real world network datasets. The results show that the communities detected by our method distribute in a smaller area compared with the traditional methods and have the similar or higher tightness on network connections. 
\end{itemize}

The rest of the paper are organized as the following. Section \ref{sec:rel} reviews the related work.  We describe the relationship between geographic location and the network structure; and propose our algorithm in section \ref{sec:alg}. We also discuss optimization and indexing methods in this section.  In section \ref{sec:exp},  we conduct experiments on both synthetic data and real world dataset. And then we give the conclusion in section \ref{sec:conc}.

\section{Related Work} \label{sec:rel}

\textbf{Community detection:} In the past decade, many algorithms have been developed to detect communities in a network. For complete discussion of various algorithms, please refer to \cite{2010community} and \cite{2011classification}. 

We only describe the most relevant work here.  Aaron \textit{et al.} provide a hierarchical clustering approach to detect communities using internal density in \cite{clauset2004finding}. The internal density is the number of edges inside a community in a network. The basic idea is to increase the ratio of the edges in communities during the hierarchical clustering process using Equation \ref{eq1}:

\vspace{-5ex}
\begin{equation}
\label{eq1}
Q=\dfrac{1}{2m}\sum_{vw}[A_{vw}-\dfrac{k_vk_w}{2m}]\delta(c_v,c_w)
\vspace{-2ex}
\end{equation}

where $A_{vw}$ is the adjacency matrix of the network and $k_v$ is the degree of node $v$.  $c_v$ represents the community of node $v$ and $\delta(c_v,c_w)$ is 1 if $c_v = c_w$.  $m$ is the number of edges in the whole network $G$. 

So the value $Q$ will be large when more edges are inside a community, which represents a good divisions of the work. To avoid the problem that the largest $Q$ value 1  will only happen when all nodes belong to the same community, the authors introduce the component ${k_vk_w}/{2m}$ in the modularity of $Q$.  ${k_vk_w}/{2m}$  is the probability of  an edge existing between nodes $v$ and $w$ if edges were randomly placed. So $Q$ will be close to zero when the network is randomly generated without community structure. Some other work is also based on modularity optimization such as \cite{blondel2008fast}, \cite{reichardt2006statistical}, and \cite{le2013fast}.

Another popular algorithm \cite{newman2004finding} is based on iteratively removing ``unimportant'' edges. The basic assumption of this method is that communities are weakly connected by a few edges. The importance of an edge, called betweenness score, is the number of shortest paths that go through that edge. The paths between different communities must go through an edge across communities so the edges across communities  will get a higher betweenness score. The edge with the highest score will be removed from the network iteratively. 

In \cite{huang2011towards}, the authors define the similarity between nodes using their degrees and the number of common neighborhood. The sum of the similarities of edges inside or outside a community was defined as internal or external similarity of a community. 

These works do not consider locations of nodes in a network.

 \begin{figure*}[ht!]
         \centering
         \subfigure[Twitter: the total variation distance is $0.315$ and the inflection distance is $4180km$]{
                 \includegraphics[ width=0.35\linewidth]{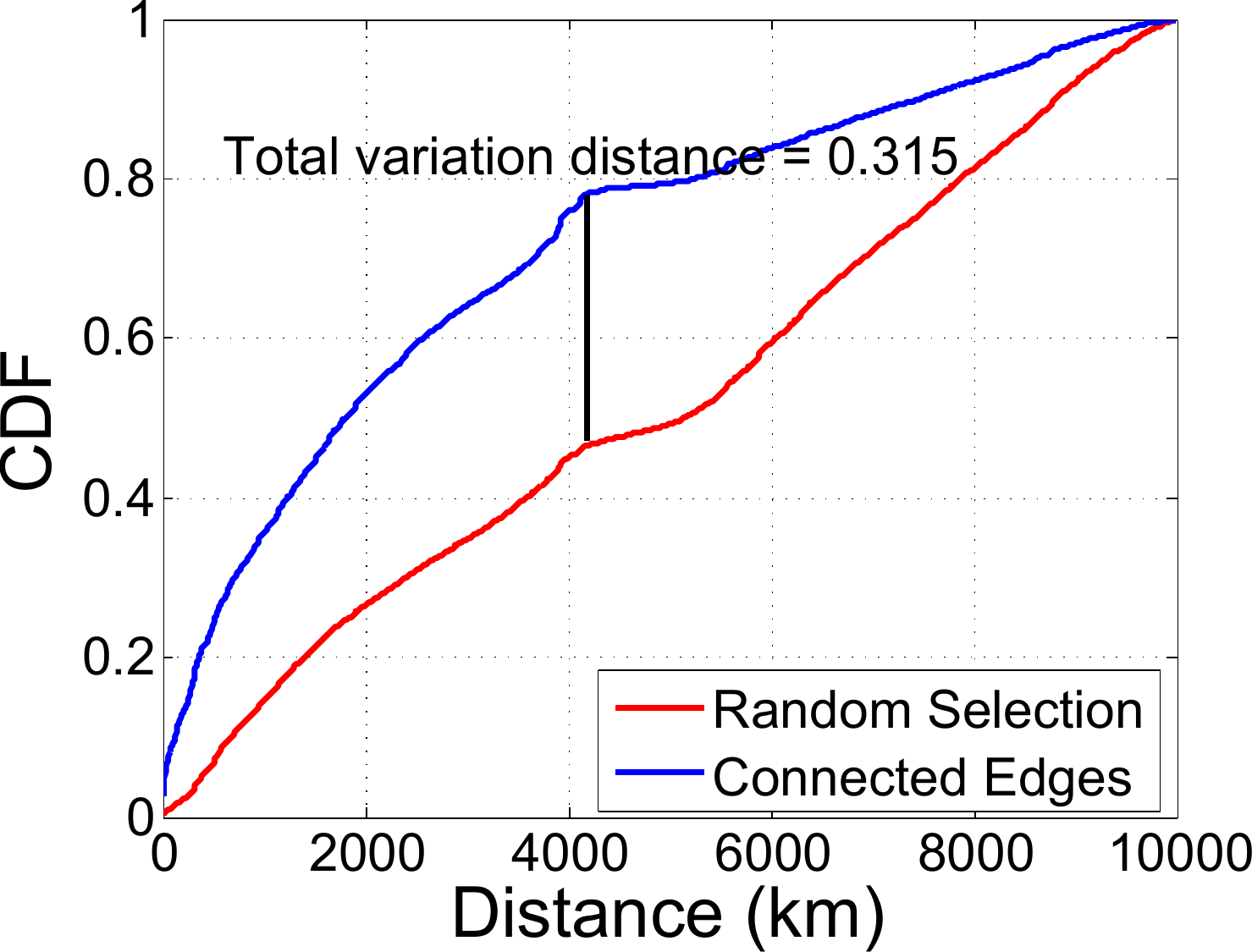}
                 \label{fig:td}
         }
         \subfigure[Gowalla: the total variation distance is $0.533$ and the inflection distance is $580km$]{
                 \includegraphics[width=0.35\linewidth]{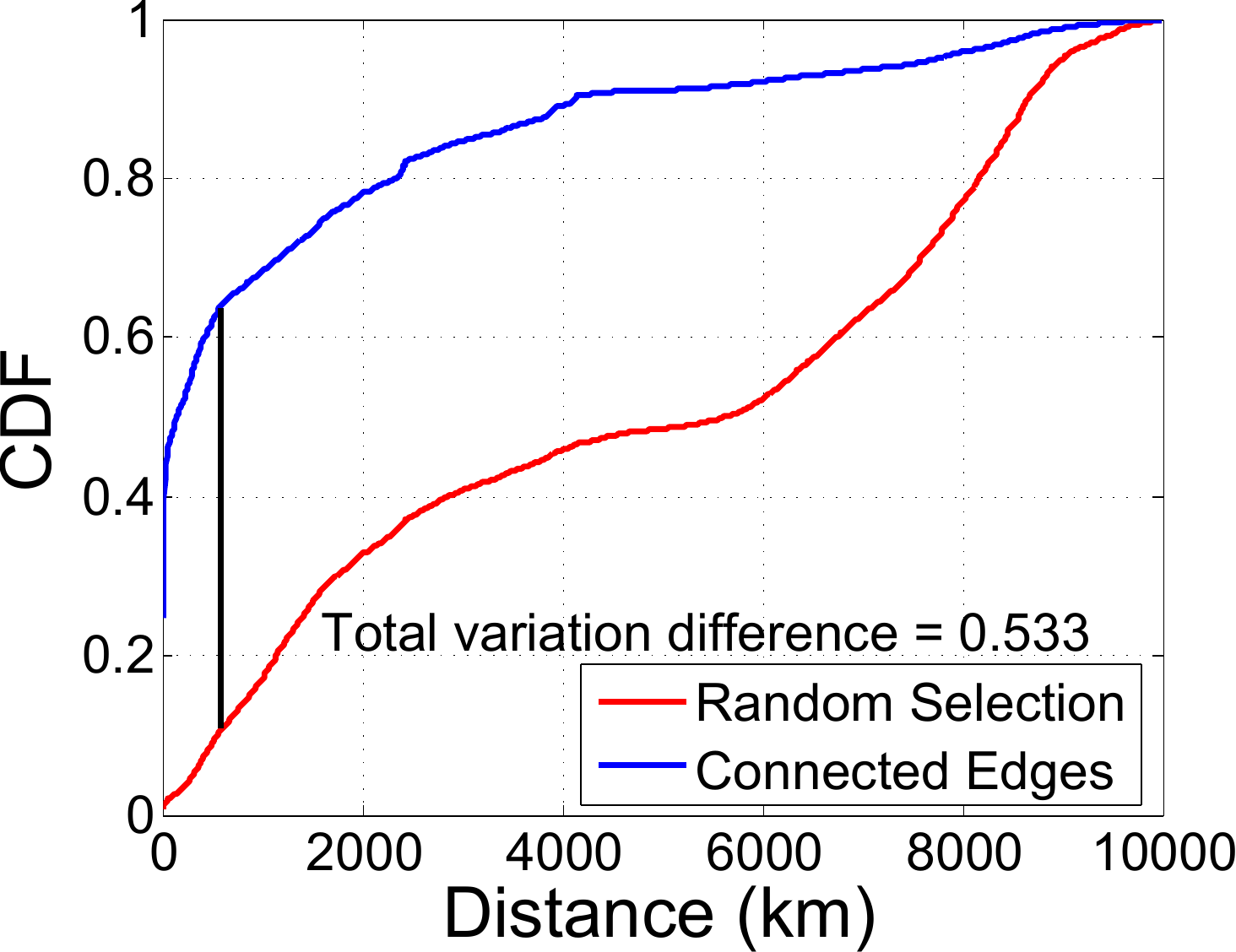}
                 \label{fig:gd}
         }
   
         \caption{The cumulative distribution function of distance between every user pair/friend pair on Twitter and Gowalla.}
         \label{fig:dis}
 \end{figure*}
 
\textbf{Geography and networks:} In the last few years, some researchers have studied the geographic constraints on real world networks. In \cite{onnela2011geographic}, the authors build a network based on the cell phone communication records. Then they study the relationship between distance and the call/text tie probability. By dividing the network into communities\cite{porter2009communities, fortunato2010community, newman2003mixing, girvan2002community}, the authors show that the geographic span of real world community is much smaller than the null community especially when the community has less than 30 people.  In \cite{scellato2010distance}, the authors define the concepts of node locality and geographic clustering coefficient. Then they show the value distribution of these two coefficients with respect to the degree of nodes. The node locality is slowly decreasing with node degree increasing. Their study shows that people tend to build connections with other nearby users. Some users have social connections only with others within a close geographic distance.

The most relevant work to ours is proposed in \cite{van2013community}.  Yves \textit{et. al.} propose a geosocial communities detection method. The authors assign each edge with a similarity score using social relationship and the Euclidean distance between their average stop locations and then run the spectral clustering algorithm.

However, the authors only built their model on a small scale application and didn't provide evidence on how the social relation is influenced by the geographic location,  which is important for using geographic information in community detection in location-tagged networks. In addition, the efficiency of the spectral clustering algorithm may be the main bottleneck when dealing with a large dataset.  In this paper, we push the location information of nodes in networks into the community detecting algorithm and design an efficient algorithm that can scale to large networks.

\section{The Algorithm} \label{sec:alg}

We denote the network as $G=(N,E,L)$, where $N$ is the set of nodes, $E$ is the edge set, and $L$ is the location set of the nodes. To determine whether the locations of nodes will help in community detection, we will analyze the locality of the network first. Then we propose our locality-based method. We follow the hierarchical clustering framework combined with the location information. A good division of the network produces communities with higher ratio of internal edges and smaller geographical scope.



\subsection{Network Locality} \label{sec:network}

As we discussed before, the formation of connections in many real world networks are influenced by the location of nodes in the network. However, some networks are more location influenced than others. So before we provide the location-based community detection algorithm, we need to analyze the influence of the location on networks to see the degree of influence. This will be helpful in determining if location based community detection is a suitable method. Here, we use network locality defined below to measure the relationship between location and connection in a network.

\vspace{-3ex}
\begin{DE}[Network Locality]
In a network $G$, we use two indexes to measure its locality: Total Variation Difference (TVD) and the Inflection Distance. Let $F(dis)$ be the cumulative distribution function (CDF) of distance between any two nodes in $G$ and $F_c(dis)$ be the CDF of the distance between connected nodes in $G$, the total variation distance is defined as:
\begin{equation}
TVD(F,F_c)=max(F_c(dis)-F(dis))
\end{equation}
and the Inflection distance is defined as the distance where $F_c(dis)-F(dis)$ achieves the maximum value.
\end{DE}
\vspace{-2ex}

From the definition, we can see that a higher value of the total variation distance indicates the network is more geographically close because connected nodes in nearby locations  have higher percentages. When the total variation difference is close to zero, the connection has little relationship with the locations of nodes. When the $TVD$ is less than zero, the connection has negative correlation with location. It is obvious that a small value of inflection distance represents a more geographic close network. 

We analyze the network locality of two real datasets: Gowalla and Twitter. Gowalla is a location-based social network and users are able to check in at ``spots'' in their local vicinity. The Gowalla dataset \cite{2011friendship} is a 196,591 users' friendship network. The check-in data were collected from February 2009 to October 2010 and each user has 32.8 check-in records on average. We use 99,563 of those users who have check-in records in our analysis. Since there is no user profile, we take the center of the $25km\times 25km$ area with the most number of check-ins as the user home location \cite{2011socio}. We also collected user profiles from Twitter, an on-line social networking and micro blogging service which allows users to follow each other; post and read ``tweets''. The data are collected from April 14 to April 28, 2013. The social network comes from \cite{2012identifying}. There are 660,000 distinct user IDs in total together with their social relations.  We obtained locations of 148,860 users through their profiles. We define the friend relation in the same way as \cite{2008social}, i.e. users $i$ and $j$ has friend relation if they follow each other. 

In Figure \ref{fig:dis}, we plot the cumulative distribution function of distance between every user pairs and friend pairs.  In the Twitter dataset, the total variation distance is $0.315$ and the inflection distance is $4,180km$. That means that the percentage of connected edges with the distance less than $4,180km$ in all the connected edges is higher than that percentage of random user pairs by $0.315$. Compared with Twitter, the Gowalla network is more close  geographically since it has a higher $TVD$, $0.533$, and a smaller inflection distance $580km$. This phenomenon illustrates that users in Gowalla tend to build friend relations with others who are geographically close to them compared with Twitter. In other words, the locations of nodes have greater influence on the network structure in Gowalla. Our experiment later also show that our method can perform better on Gowalla than Twitter network for this reason. 

In practice, the total variation difference is more helpful to measure how the network structure is influenced by location. We suggest applying our method on the networks with the total variation difference larger than $0.25$. 
  
\subsection{Connection Locality}

To take location into account in community detection, first we define the concept of connection locality to qualify the graphic closeness between nodes.
\vspace{-2ex}
\begin{DE}[Connection Locality]
 Let $dis_{vw}$ be the geographic distance between nodes $v$ and $w$. Let $\sigma$ be the average distance between all user pairs. The connection locality can be defined as:
\begin{equation}
L_{vw}=exp(-dis_{vw}/\sigma)
\end{equation}
\end{DE}

So connection locality will achieve a high value when the two nodes are close. Since our goal is to detect communities with both geographic closeness and network tightness, we measure the geographic and network closeness of the communities using the following equation:

\vspace{-3ex}
\begin{equation}
C_G = \dfrac{1}{\sum_{vw}A_{vw}L_{vw}}\sum_{vw}A_{vw}L_{vw}\delta(c_v,c_w)
\vspace{-1ex}
\end{equation}

We can see that this method is equivalent to assign each edge in network $G$ with the locality as weight.  Inspired by the method in \cite{clauset2004finding}, we introduce the expected value of $C_G$ to avoid the situation that the largest value of $C_G$ will be achieved when all the nodes belong to the same community.

\begin{figure}[ht!]
        \centering
        \includegraphics[width=0.95\linewidth]{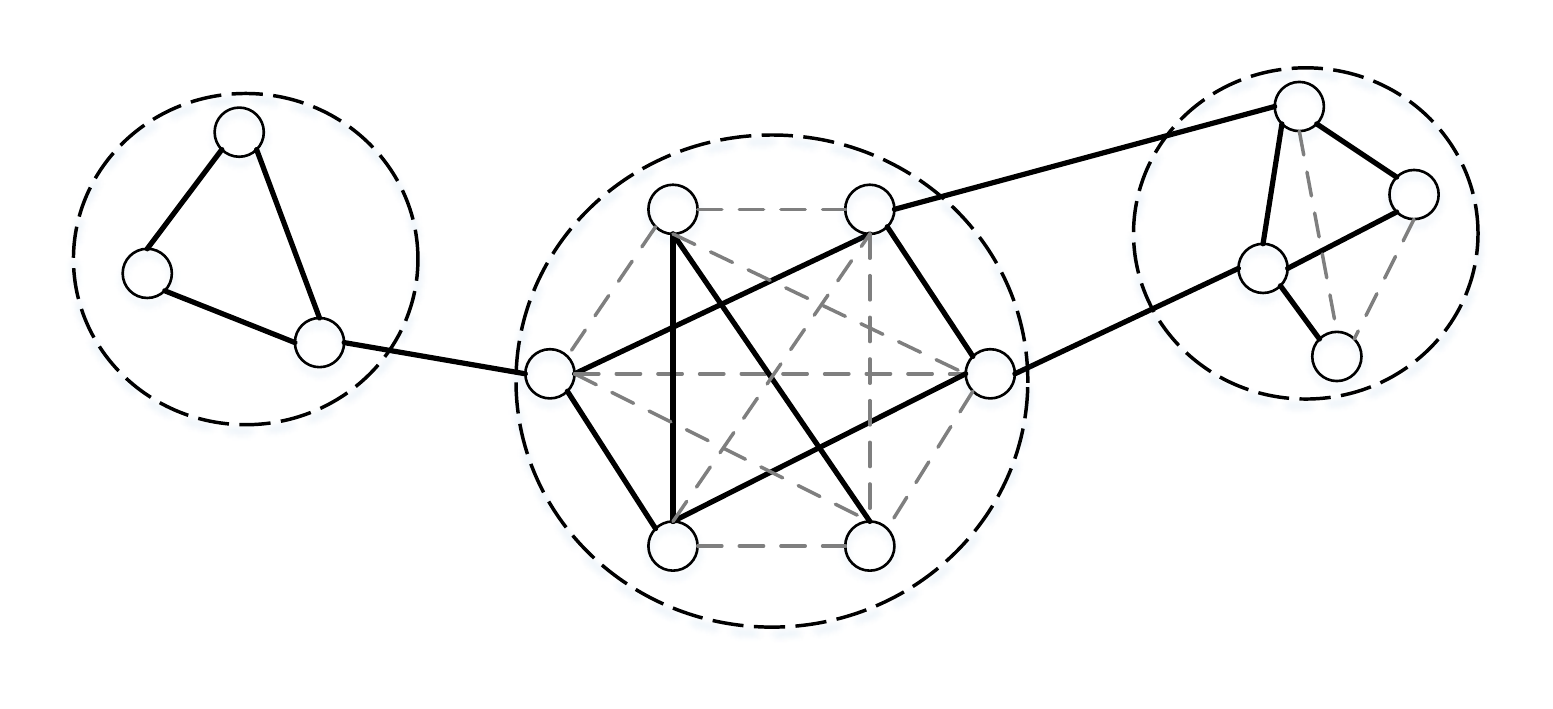}
        \caption{In this example, each dashed circle represents a community. The solid lines are connections between nodes.}
        \label{fig:complete}
\end{figure}

The expected value of $C_G$ is obtained from a random connection network. For a given network $G$, the location of the nodes and their degrees are fixed. In a random connection network, the probability of an edge existing between a node pair is $k_vk_w/2m$. Since we already know the locations of the nodes, the community locality $l_{vw}$ between a node pair $v$ and $w$ is also known. So the expect value for each edge is $l_{vw}k_vk_w/2m$ and the expect value of $C_G$ is the sum of the expect value of all the edges:

\vspace{-3ex}
\begin{equation}
P_G = \dfrac{1}{\sum_{vw}A_{vw}L_{vw}}\sum_{vw}\dfrac{k_vk_w}{2m}L_{vw}\delta(c_v,c_w)
\vspace{-1ex}
\end{equation}

In Figure \ref{fig:complete}, there are three communities denoted by the dashed circles. The solid lines are the edges in the network. We use dashed lines to complement each community as a complete graph. Given the locations of all nodes,  the community localities can be easily calculated. The probability of  an edge existing between the bottom two nodes in the left community is $\frac{2 \times 2}{13}$. 

Let $\omega = \sum_{vw}A_{vw}L_{vw}$, we define the modularity Q as: 

\vspace{-3ex}
\begin{equation}\label{eq:b}
Q =  \dfrac{1}{\omega}\sum_{vw}[A_{vw}L_{vw}-\dfrac{k_vk_w}{2m}L_{vw}]\delta(c_v,c_w)
\vspace{-1ex}
\end{equation}

The community locality between each node pair is fixed. If the network is not built based on locations, community locality will have no influence on $Q$ and the value of the modularity $Q$ will be close to zero. When the network has a good divisions, which means the communities are close on both geographic distance and network structure, the modularity $Q$ will achieve a higher value.

\subsection{Node Similarity}

In this section, we will discuss how to enhance the influence of network structure in the detection process. Here we define the node similarity between nodes pair by the common neighbors and their degrees:

\vspace{-2ex}
\begin{DE}[Node Similarity]
Let $\Gamma_v$ be the set of neighbors of vertex $v$. The similarity of two nodes is calculated by their common neighbors and their degrees as:
\begin{equation}
S_{vw} = \dfrac{|\Gamma_v \cap \Gamma_w|}{\sqrt{|\Gamma_v||\Gamma_w|}}
\end{equation}
\end{DE}
\vspace{-2ex}
\begin{figure*}[ht!]
        \centering
        \subfigure[1-degree node pairs]{
                \includegraphics[ width=0.41 \linewidth]{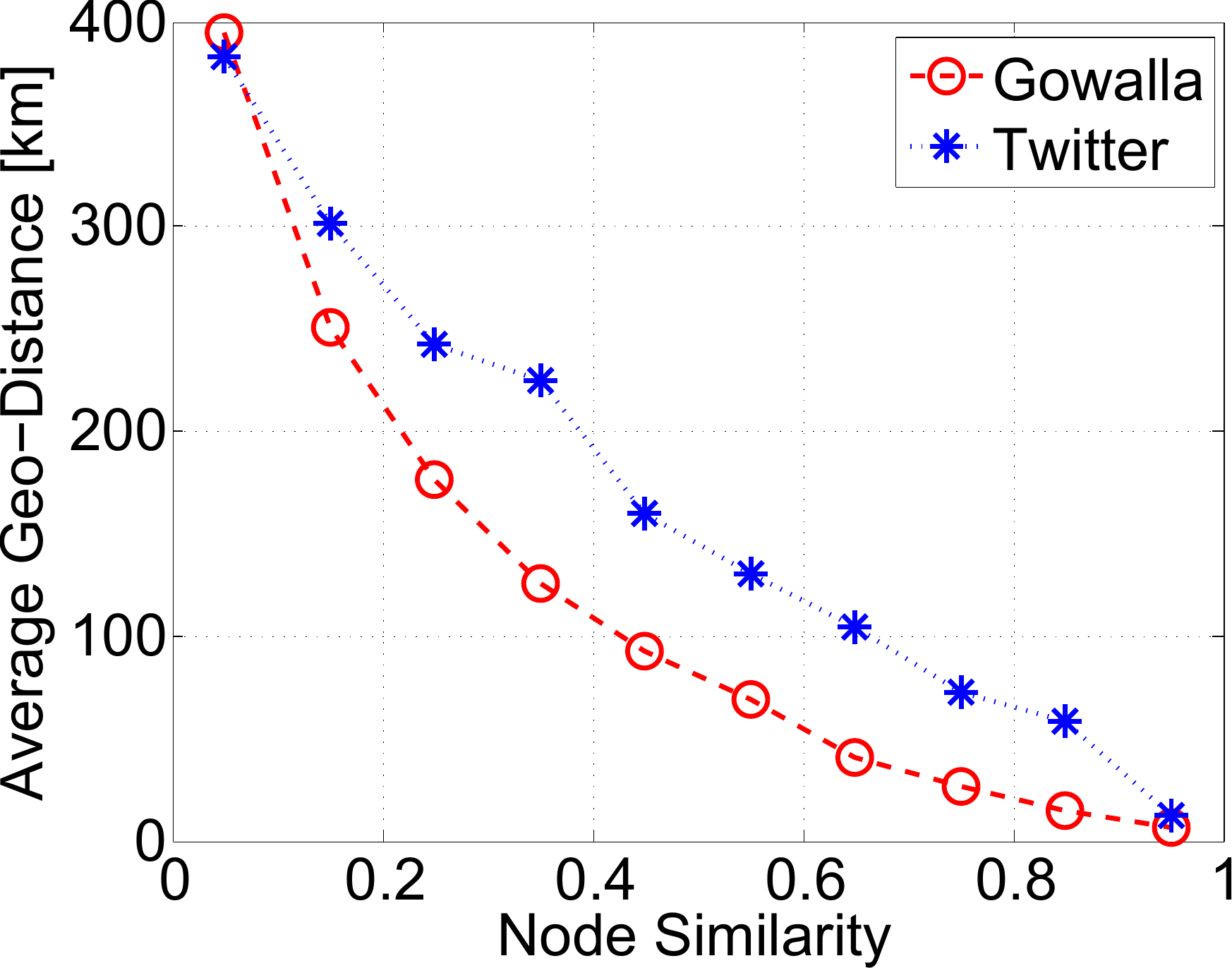}
                \label{fig:ave}
        }
        \subfigure[1- and 2-degree node pairs]{
                        \includegraphics[ width=0.4\linewidth]{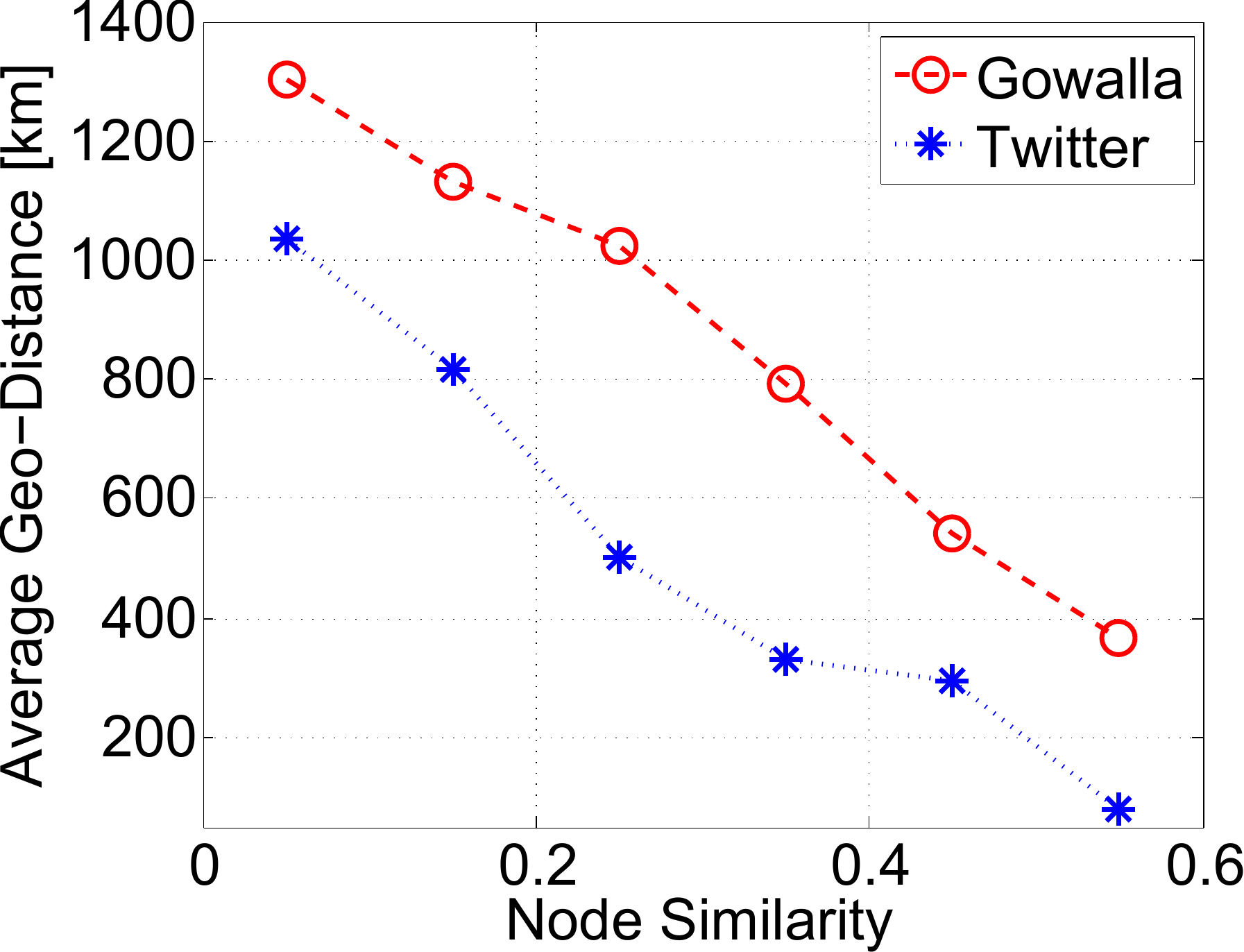}
                        \label{fig:coall}
                }

        \caption{The average geographic distance under different values of node similarity.}
        \label{fig:cosine1}
\end{figure*}

First we study the relationship between node similarity and the geographic distance between node pairs in real world networks. From Figure \ref{fig:ave} we can see that with the increasing of the value of node similarity, the average geographic distance has a significant decrease. Then we extend the investigation of node similarity to all 1- and 2-degree friend pairs. From Figure \ref{fig:coall} we can see a similar tendency between node similarity and average geographic distance as that on 1-degree friends. But the average distances are much longer than the 1-degree friends especially on the Gowalla data set.

To apply the node similarity in our modularity, we also need to calculate the expect value under random connection network. In the random case, when we calculate the expect value of $S_{vw}$, we need to know the probability of an edge existing between node $v$ or $w$ and any other node $i$. Assuming node $i$ has $k_i$ neighbors, the probability of node $i$ is connected to $v$ ($w$) is $k_vk_i/2m$ ($k_wk_i/2m$). Since the probability of connection is independent, so the probability of node $i$ connected to both $v$ and $w$ is $\frac{k_vk_i}{2m}\frac{k_wk_i}{2m}$. The expected value of $S_{vw}$ is the sum of the probabilities of both $v$ and $w$ connected to any other node $i$:

\vspace{-3ex}
\begin{equation}\label{eq:sc}
\begin{split}
S_{vw} & = \dfrac{|\Gamma_v \cap \Gamma_w|}{\sqrt{|\Gamma_v||\Gamma_w|}} \\
 & =  \sum_{i \neq v \& i \neq w} (k_vk_i/2m)(k_wk_i/2m) / \sqrt{k_vk_w}\\
  &  = \sqrt{k_vk_w}\sum_{i \neq v \& i \neq w} k_i^2 / 4m^2 \\
\end{split}
\end{equation}

In practice, we use $\tau = \sum_i k_i^2 / 4m^2$ instead of $\sum_{i \neq v \& i \neq w} k_i^2 / 4m^2$ because they have similar value on larger networks. 

We then revise $\omega$ as $\sum_{vw}A_{vw}S_{vw}L_{vw}$, and the new modularity $Q^s$ is defined as:
\vspace{0ex}
\begin{equation}\label{eq:f}
Q^s=\dfrac{1}{2\omega}\sum_{vw}[A_{vw}S_{vw}L_{vw}-L(v,w)\dfrac{k_vk_w}{2m}\tau\sqrt{k_vk_w}]\delta(c_v,c_w)
   \vspace{-2ex}
\end{equation}

In this paper, we only consider the node similarity between connected nodes for the following reasons: (1) Relation of 2-degree neighbors (the node pairs which are connected but share at least one common neighbors) introduce many new connections. The number of 2-degree neighbors is much more than directly connected neighbors and will significantly increase the computation complexity. (2) The influence of 2-degree neighbors is much smaller than directly connected ones. Based on our investigation, the average distance between 2-degree neighbors are three to times times longer than directly connected neighbors even when they have the same node similarity.

\subsection{Optimization and Complexity}
In this section, we will discuss how to implement the algorithm efficiently with optimization and indexing. The algorithm is based on the hierarchical clustering method with greedy strategy. At first, each node is a community.  In each step, two communities whose combination increases the value of the modularity $Q$ most are combined. In \cite{clauset2004finding}, the authors provide an efficient method to implement their model. They maintain and update a matrix $\Delta Q_{ij}$ which records the change of $Q$ after combing the communities $i$ and $j$. When Equation \ref{eq:b} is used as $Q$ value, we can implement the model in a similar way. When Equation \ref{eq:f} is used as the modularity, we discuss the optimization here.

\vspace{-2ex}
\begin{equation}\label{eq:f1}
\begin{split}
Q&=\dfrac{1}{2\omega}\sum_{vw}A_{vw}S_{vw}L_{vw}\delta(c_v,c_w) \\
 & - \dfrac{1}{2\omega}\sum_{vw}L(v,w)\dfrac{k_vk_w}{2m}\tau\sqrt{k_vk_w}\delta(c_v,c_w)\\
\end{split}
\end{equation}

We can rewrite the modularity in Equation \ref{eq:f} into Equation \ref{eq:f1}. By analyzing the modularity, we can see that after we combine two communities $i$ and $j$, the change of $Q^s$ includes two parts: 1) the connections between these two communities will increase the value of $Q^s$ (the first part in Equation \ref{eq:f1}), the value equals to:

\vspace{-3ex}
\begin{equation}\label{eq:p1}
\Delta Q_1 = \dfrac{1}{2\omega}\sum_{vw}A_{vw}S_{vw}L_{vw}\delta(c_v,i)\delta(c_w,j)
\end{equation}

and 2) the value generated by node pairs from communities $i$ and $j$ and this value equals to: 

\vspace{-3ex}
\begin{equation}\label{eq:p2}
\Delta Q_2 = - \dfrac{1}{2\omega}\sum_{vw}L(v,w)\dfrac{k_vk_w}{2m}\tau\sqrt{k_vk_w}\delta(c_v,i)\delta(c_w,j)
\end{equation}

So the $\Delta Q_{ij}$ equals to $\Delta Q_1 + \Delta Q_2$. Since we will combine two communities with the largest $\Delta Q_{ij}$, we only need to keep values in Equations \ref{eq:p1} and \ref{eq:p2}. Now we only need to solve two problems, how to initialize the $\Delta Q_{ij}$ and how to update it after we combine two communities.

The combination of two disconnected communities will not increase the value of $Q$, so we only keep the $\Delta Q_{ij}$ if there is at least one edge between them. At first, every node is a community and the $\Delta Q$ between each connected node pair is: 

\vspace{-3ex}
\begin{equation}\label{eq:ini}
L_{ij}[\dfrac{S_{ij}}{2 \omega} - \dfrac{\tau(k_ik_j)^{1.5}}{4\omega m}]
\end{equation}

After we combine communities $i$ and $j$, we need to update all the communities $k$ which are connected to $i$ or $j$. We use $(ij)$ to denote the community generated by combining $i$ and $j$ and use $\Delta Q_{k, (ij)}$ to denote the new $\Delta Q$ value between $k$ and $(ij)$. If the community $k$ is connected to both $i$ and $j$, we can get the new $\Delta Q_{k, (ij)}$ by $\Delta Q_{ik} + \Delta Q_{jk}$. If $k$ is only connected to one of them, e.g. $i$, we do not have $Q_{jk}$ since they are disconnected. So we need to calculate it. We have already known that $\Delta Q$ is the sum of Equation \ref{eq:p1} and \ref{eq:p2}. The $\Delta Q_1$ will be zero since there is no edge between $j$ and $k$. So we can have the $\Delta Q_{jk}$ as:

\vspace{-3ex}
\begin{equation}\label{eq:jk}
\Delta Q_{jk} = - \dfrac{1}{2\omega}\sum_{vw}\tau L(v,w)\dfrac{(k_vk_w)^{1.5}}{2m}\delta(c_v,j)\delta(c_w,k)
\end{equation}

And then we can update the $\Delta Q_{k, (ij)}$ by:

\vspace{-3ex}
\begin{equation}\label{eq:update}
\Delta Q_{k, (ij)}=\Delta Q_{ik} + \Delta Q_{jk}
\end{equation}

\begin{algorithm}
\fontsize{9}{9}\selectfont
\begin{algorithmic}[1]
\STATE \textbf{Input:} Network $G = (N, E, l)$
\STATE \textbf{Output:} Communities in $G$
\STATE Assign each node a community label from 1 to n
\STATE Initialize the $\Delta Q_{ij}$ as Eq.\ref{eq:ini}
\STATE Find the maximum $\Delta Q_{ij}$, $max\Delta Q$
\WHILE{$max\Delta Q$ > 0}
	\STATE Update $\Delta Q_{k, (ij)}$ of all the communities $k$ connect to $i$ or $j$ by Eq.\ref{eq:jk} and Eq.\ref{eq:update}
	\STATE Update the community label in community $i$ as $j$
\ENDWHILE
\STATE Return node list with the community label
\end{algorithmic}
\caption{Detecting communities from location-tagged network}
\label{alg:detection}
\end{algorithm}

The Algorithm \ref{alg:detection} describes the frame work of all the process. We will stop the hierarchical clustering process when the modularity $Q$ achieve its maximum value, which means that the largest $\Delta Q_{ij}$ is less than zero.

We store each row of the $\Delta Q_{ij}$ and the node list in different communities in a balanced binary trees. When we update a $\Delta Q_{k, (ij)}$, the worst case is that we need to calculate the $\Delta Q_{jk}$ by Equation \ref{eq:jk} and the complexity is $O(|j||k|log n)$, where $|j|$ represents the number of nodes in community $j$. For each combination of two communities, the worst case is that all the nodes connected to all the communities. Assume that the depth of the hierarchical clustering is $d$ and the number of nodes in a community is $c_n$,  the complexity is $O(mdc_n^2logn)$.

\section{Experiment} \label{sec:exp}
In this section, we test our method on synthetic networks and two real world social network datasets described before: Twitter and Gowalla. We use three different measurements to evaluate the results:

\begin{DE}[Geographic Span]
The geographic span of a community $c$ is defined as the average distance of the nodes in $c$ to the centroid $(\bar{x},\bar{y})$ of all the nodes in the community:
\begin{equation}
S(c) = \frac{1}{|c|}\sum_v \sqrt{(x_v - \bar{x})^2 + (y_v-\bar{y})^2} \delta(c_v,c)
\end{equation}
\end{DE}

\begin{DE}[Average Internal Degree]
The internal degree of a node $v$ is the number of its neighbors in the same community. The average internal degree of a community $c$ is the average value of the internal degrees of all the nodes in $c$ and it can be represented as:
\begin{equation}
A(c) = \frac{1}{|c|}\sum_{vw} \delta(c_v,c)\delta(c_w,c)
\end{equation}
\end{DE}

The last measurement is the detection accuracy. Since we do not have a class label of the real world datasets, we only apply this on the synthetic networks. We implemented four community detection methods in our experiments: 1) Randomly select nodes as community (Random). 2) The method proposed in \cite{clauset2004finding} (Clauset's Method). 3) The method discussed in section \ref{sec:alg} using  Equation \ref{eq:b} as the modularity $Q$ (Connection Locality). 4) The method discussed in  section \ref{sec:alg} with Equation \ref{eq:f} as the modularity (Node Similarity).

\begin{table*}[htb]
\caption{The accuracy of different community detection methods}
\centering
\begin{tabular}{c|c|c|c}
\toprule[1pt]
$\Omega$ & Clauset's Method & Connection Locality & Node Similarity  \\ 
\hline 
1 & 16.24  & 16.63 &  18.38 \\ 
3 & 16.48 & 22.82 & 24.63 \\ 
5 &  17.72 & 22.40 & 28.77 \\
10 & 22.16  & 25.14 & 26.60 \\
30 & 32.84  & 19.76 & 24.42 \\
$+\infty$ &  36.04 & 19.20 & 19.76 \\
\bottomrule[1pt]
\end{tabular} 
\label{tab:acc}
\end{table*}

\begin{figure*}[ht!]
        \centering
        \subfigure[$\Omega = 3$]{
                \includegraphics[ width=0.3\linewidth]{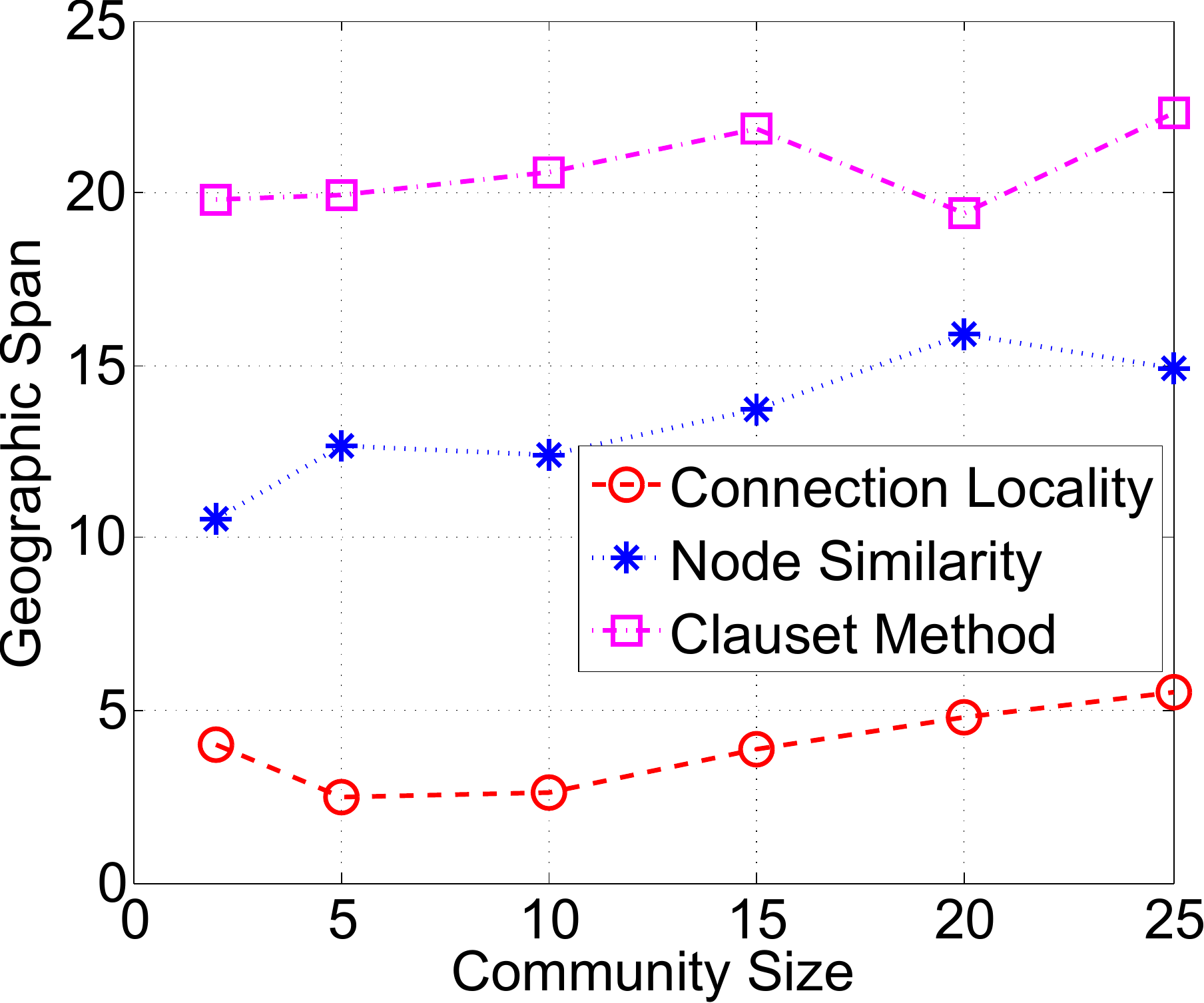}
                \label{fig:1}
        }
        \subfigure[$\Omega = 10$]{
                        \includegraphics[ width=0.3\linewidth]{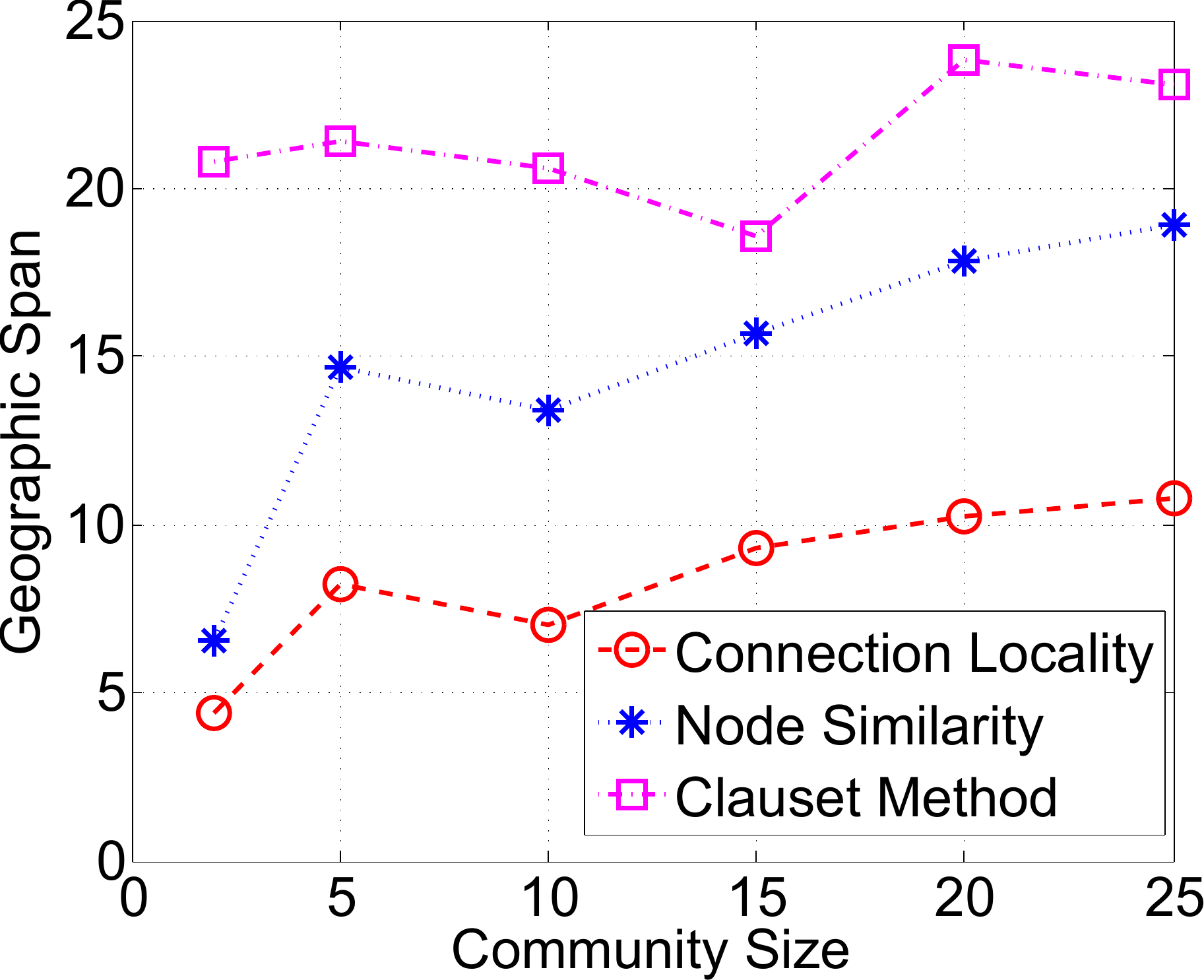}
                        \label{fig:2}
                }
         \subfigure[$\Omega = +\infty$]{
                         \includegraphics[ width=0.3\linewidth]{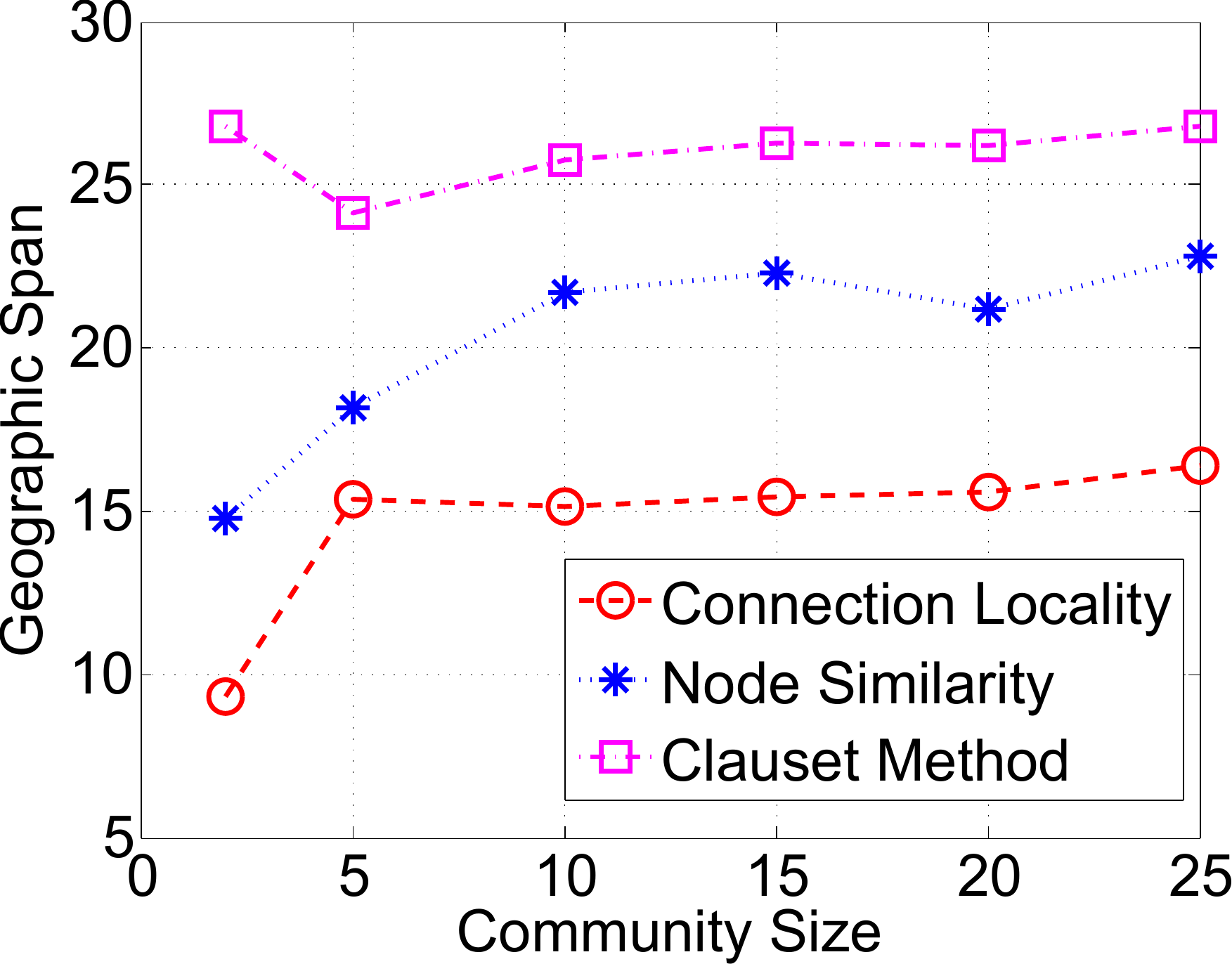}
                         \label{fig:3}
                 }
        \subfigure[$\Omega = 3$]{
                        \includegraphics[ width=0.3\linewidth]{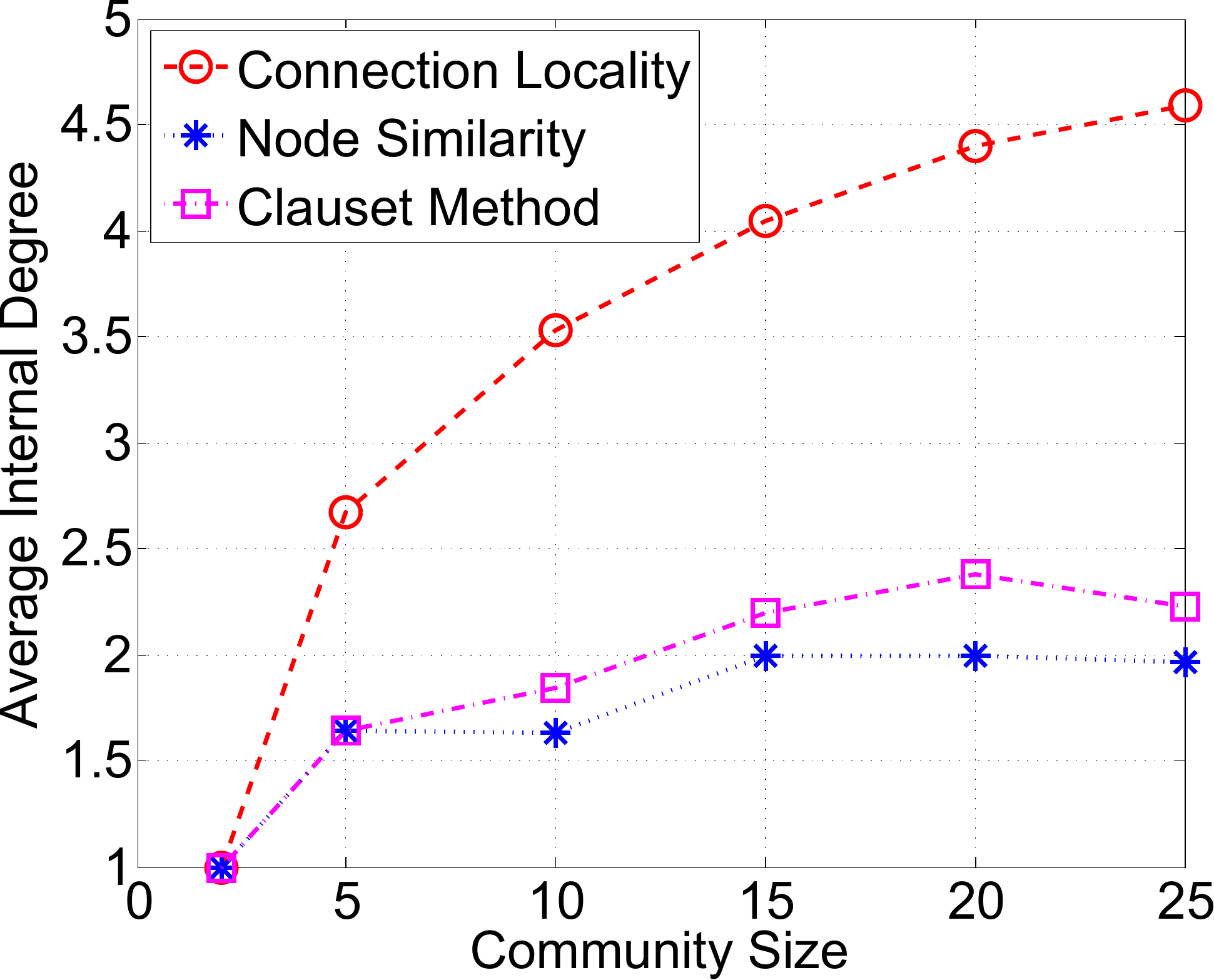}
                        \label{fig:4}
                }
        \subfigure[$\Omega = 10$]{
                        \includegraphics[ width=0.3\linewidth]{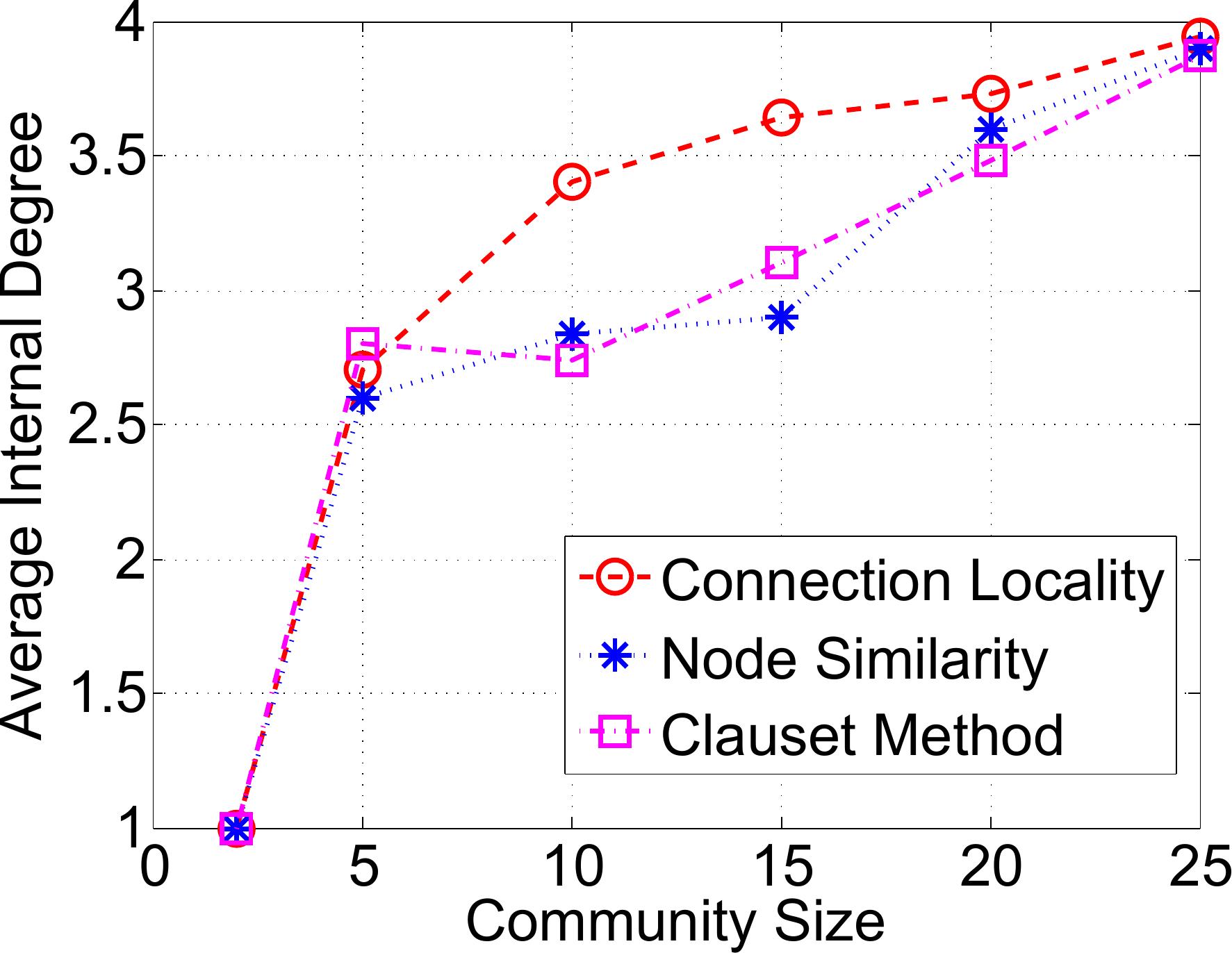}
                        \label{fig:5}
                }
        \subfigure[$\Omega = +\infty$]{
                        \includegraphics[ width=0.3\linewidth]{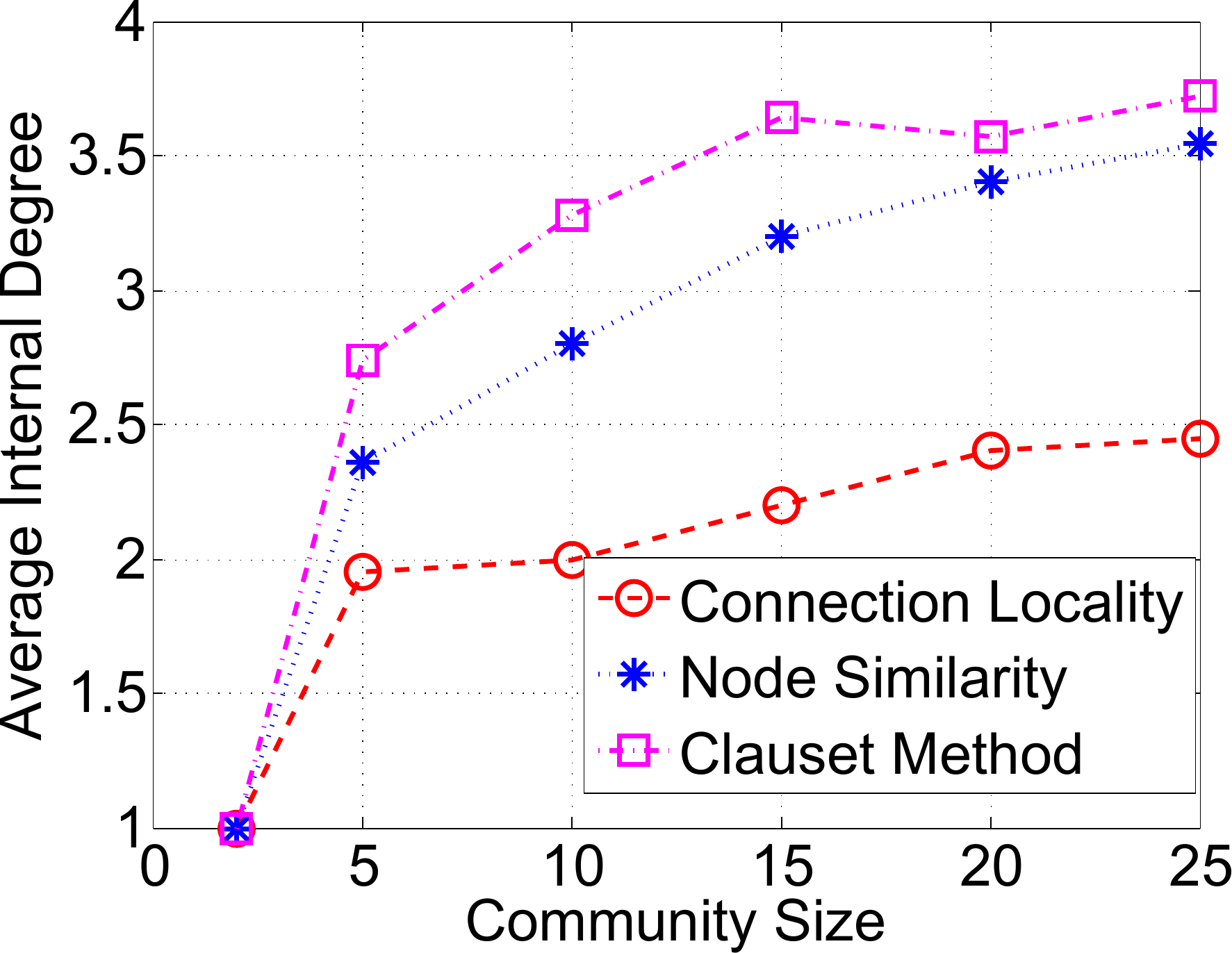}
                        \label{fig:6}
                }
       
        \caption{The geographic span and average internal degree of the synthetic network under different values of $\Omega$.}
        \label{fig:synthetic}
\end{figure*}

\subsection{Tests on Synthetic Networks}

First we test the methods on the generated networks because a synthetic datasets allow for better parameter control. We analyze the results using the three measurements discussed above. When generating the dataset, we control the influence of geographic distance on building connection between two nodes in order to see the how the geographic feature affect the detection methods.

We generate the networks on a $50\times 50$ grid.  There are 2,500 nodes in total in the network. For each node, we randomly assign a community label to it. There are 10 different community labels in the network. We generate the probability of an edge existing between node $v$ and $w$ as:

\vspace{-3ex}
\begin{equation}
p_e = \alpha p_c e^{-dis_{vw}/\Omega}
\end{equation}

The value of $p_c$ depends on whether the node $v$ and $w$ have the same community label. If their community labels are the same, $p_s$ is set to $0.5$ and if not, $p_c$ is set to $0.1$. So the edges have a higher probability of occurrence between the nodes with the same community label. The component  $e^{-dis_{vw}/\Omega}$ is used to control the influence of the locations of nodes. When we set a large enough value to $\Omega$, the value of $e^{-dis_{vw}/\Omega}$ is close to 1 and the probability is almost not influenced by $dis_{vw}$. So the network structure is not influenced by the locations of nodes. On the contrary, if the value of $\Omega$ is small, the value of $e^{-dis_{vw}/\Omega}$ will be greatly influenced by the distance between $v$ and $w$. In that case, only the nearby neighbors with the same community label will have a high probability of connecting. The parameter of $\alpha$ is used to control the average degrees of nodes in the network. In the following experiment, we make the number of average degree around 15 by adjust the value of $\alpha$.

Table \ref{tab:acc} shows the accuracy of different algorithms on different generated networks. Since the largest distance in the network is only 70, when we set the $\Omega$ larger than 10, the probability of connecting is not very sensitive to the distance. We can see that when the $\Omega$ is less than 10, which means the building of connections is greatly influenced by the location of nodes, our two methods can achieve a similar or higher accuracy than the Clauset's method. With the increasing of the value of $\Omega$, when the location has little or no influence on the network structure, the accuracy of Clauset's method performance better than our methods. So we recommend to evaluate the influence of geographic information first as described in section \ref{sec:network} before applying our methods on a network.

Figure \ref{fig:synthetic} shows how the three methods perform on different synthetic networks. From Figure \ref{fig:1} to \ref{fig:3} we can see for all levels of influence ($\Omega$) that the geographic location on network structure, the connection locality have the smallest geographic span.  The geographic span of the node similarity method is smaller than the Clauset's method. Figure \ref{fig:4} to \ref{fig:6} show the average internal degree of the three methods. When  $\Omega$ is 3, where  the location of nodes will have the greatest influence on the network structure, the connection locality method has a higher value of internal degree. This illustrates that this method is suitable to deal with the highly geographically influenced networks. When the value of $\Omega$ increases to 10, the average internal degree of these three methods is similar. But when we set the $\Omega$ as infinity, the connection locality and node similarity method perform worse than Clauset's method. 
 
\subsection{Twitter Network}

\begin{figure*}[ht!]
        \centering
        \subfigure[The geographic span of different size of communities detected by different methods]{
                \includegraphics[ width=0.4\linewidth]{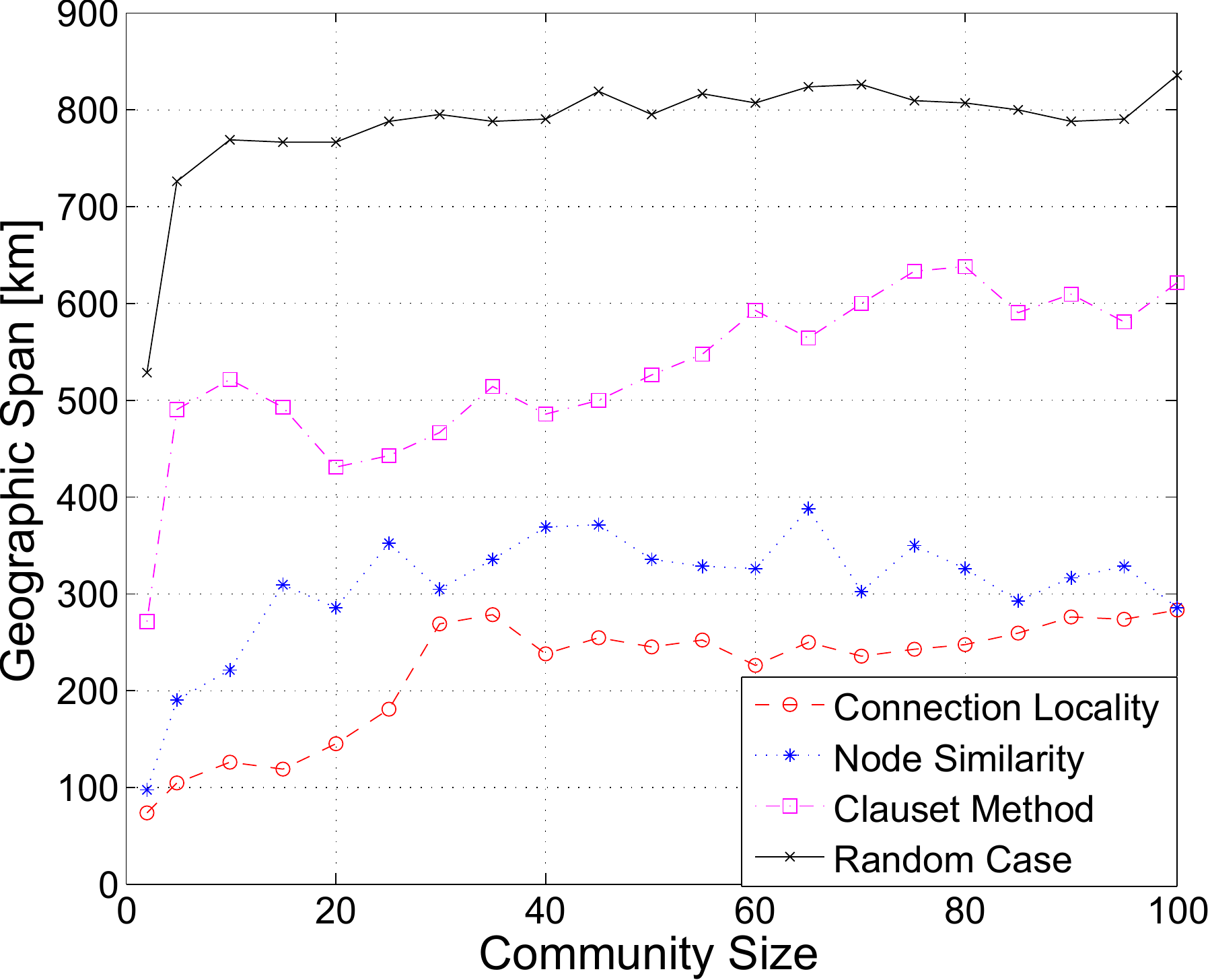}
                \label{fig:twitterdis}
        }
        \subfigure[The average internal degree of different size of communities detected by different methods]{
                \includegraphics[width=0.41\linewidth]{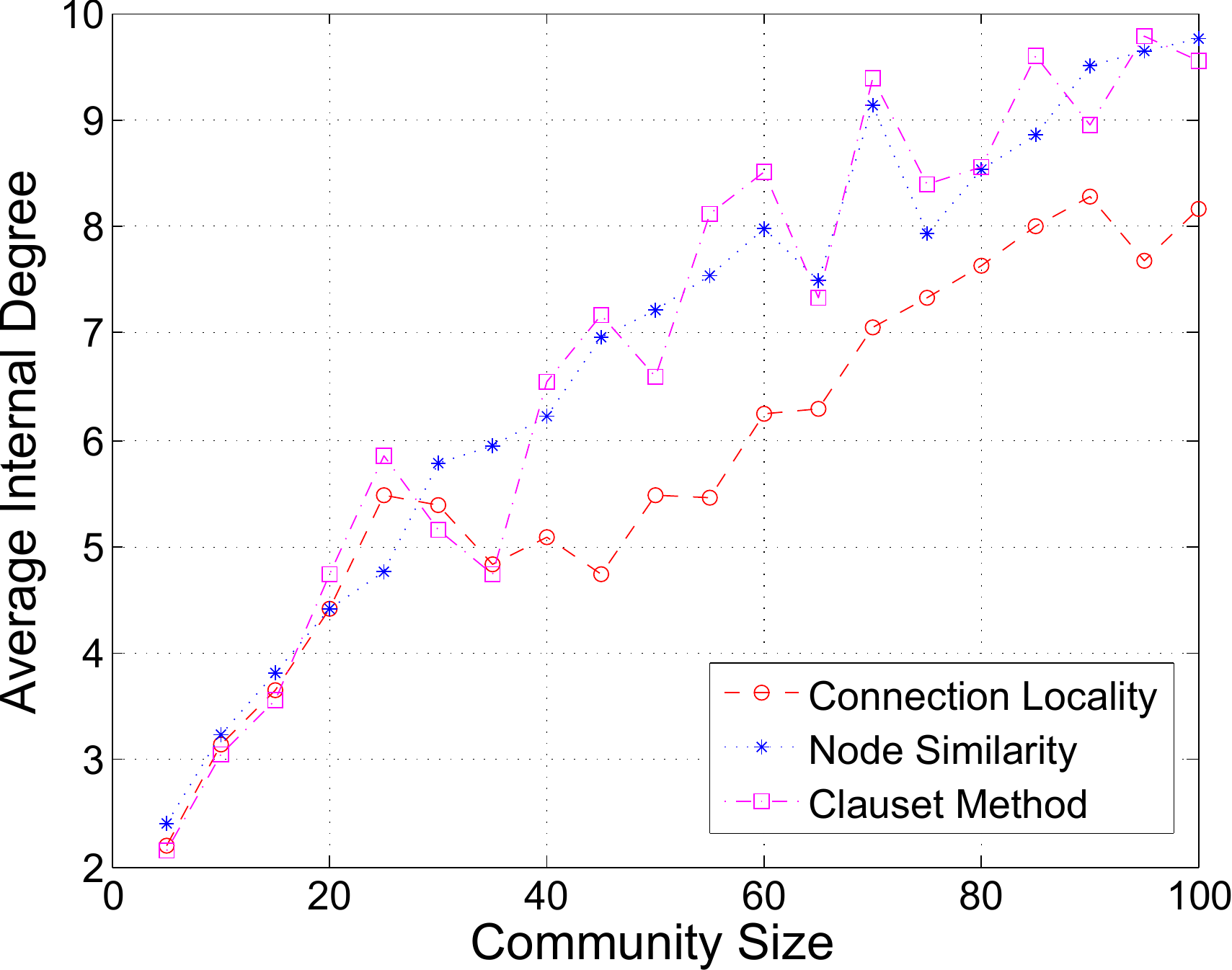}
                \label{fig:twitterdegree}
        }
  
        \caption{Analyzing the community detection results of different methods on the Twitter Network.}
        \label{fig:twitter}
\end{figure*}

\begin{figure*}[ht!]
        \centering
        \subfigure[The geographic span of different size of communities detected by different methods]{
                \includegraphics[ width=0.41\linewidth]{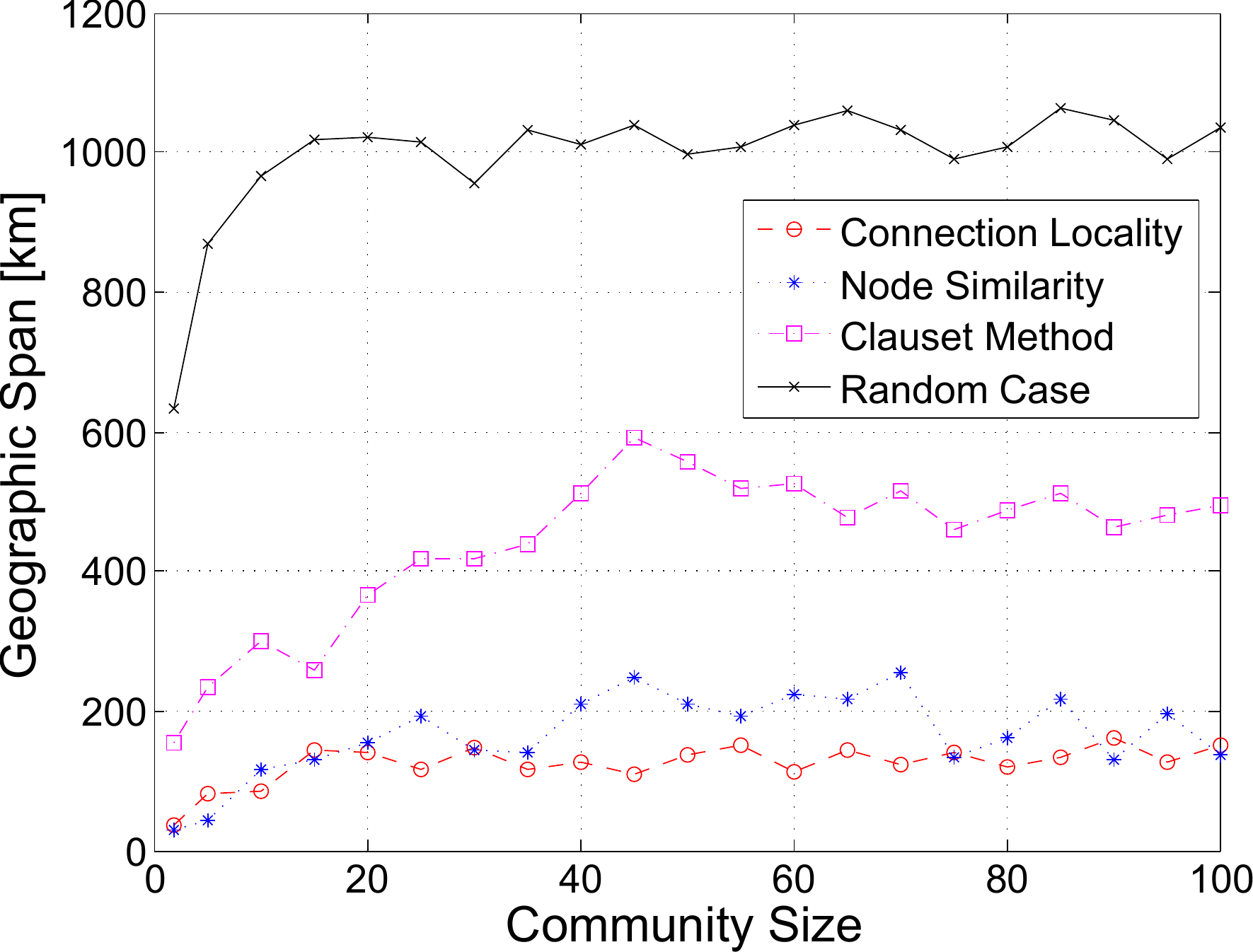}
                \label{fig:gowalladis}
        }
        \subfigure[The average internal degree of different size of communities detected by different methods]{
                \includegraphics[width=0.40\linewidth]{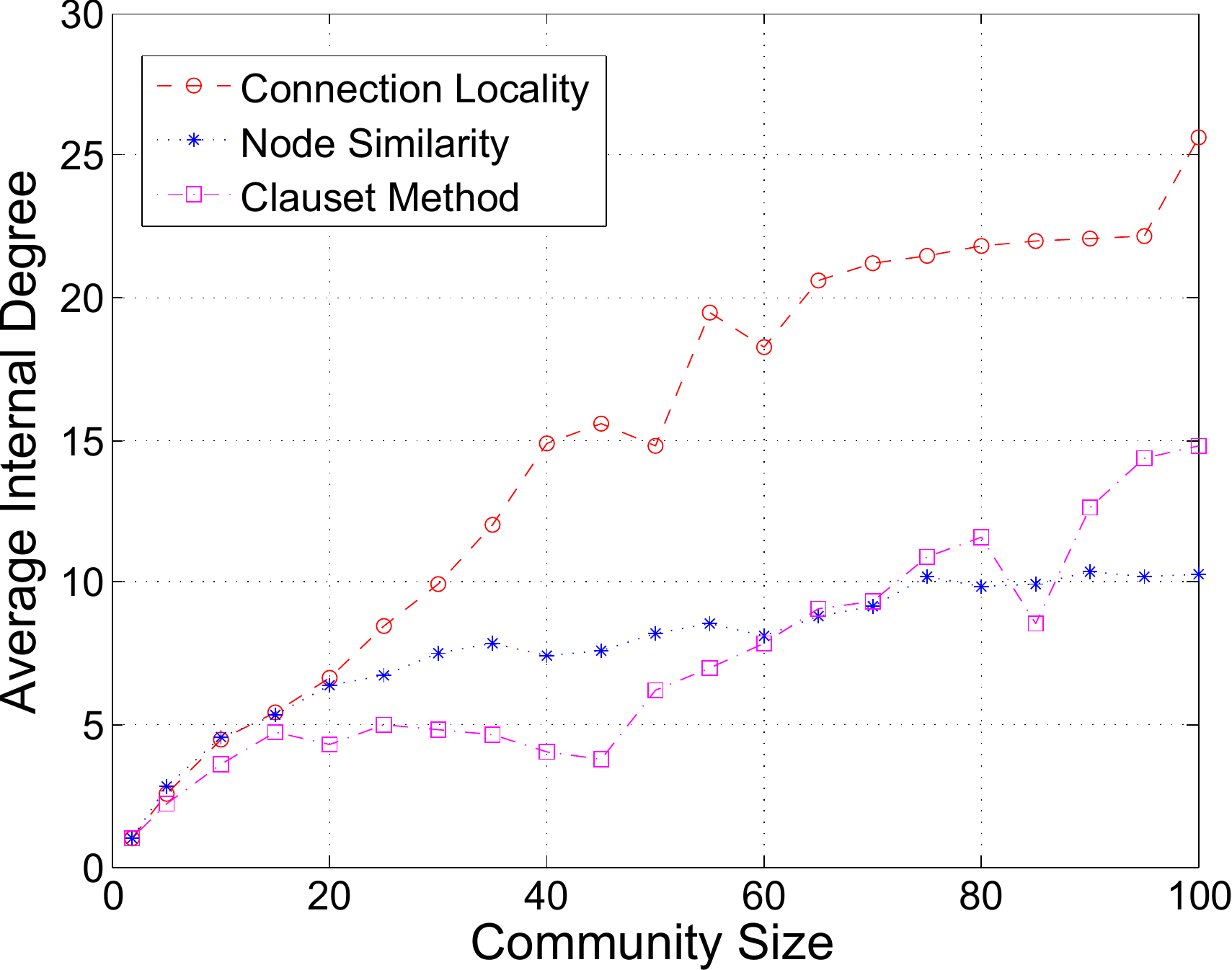}
                \label{fig:gowalladegree}
        }
  
        \caption{Analyzing the community detection results of different methods on the Gowalla Network.}
        \label{fig:gowalla}
\end{figure*}

In the real world, the factors which can influence the network structure can be very complex. We now test the algorithms on the networks generated by some real world applications. The first example is the Twitter network. We have introduced the details of this network in Section \ref{sec:network}. Since we do not have a community label for the real world dataset, we only apply the geographic span and the average internal degree of the communities to evaluate the detection results.

In Figure \ref{fig:twitterdis}, we demonstrate the geographic span of different sizes (number of nodes in the community) of communities. From this figure we can see that under the random case, the geographic span is much larger and increases quickly to $800$ kilometers. The communities detected by Clauset's method has a smaller geographic span. It begin with 280 kilometers when the community size is 2 but increases quickly when the community size become larger. Finally, the geographic span fluctuates between 500 to 600 kilometers. The two methods proposed in this paper have the best performance on controlling the geographic span on communities. Although the geographic span increases quickly when the community size becomes larger, these two method can keep the span much smaller than Clauset's method and the random case, especially for the method with the Equation \ref{eq:b} as the modularity. The geographic spans in different sizes of communities are only half of Clauset's method.

The Figure \ref{fig:twitterdegree} shows the average internal degrees of different sizes of communities. This measurement evaluates the detection result by the network structure only. From the definition we know that if a community have a higher internal degree, that means the connections inside the community is tighter. From the figure we can see that with the increasing of the community size, the average internal degree also becomes larger, which means nodes have more neighbors in the same community with them. The Clauset's method and one of our method, which use $Q^s$ as the modularity, have a similar performance. The connection locality method has a smaller average internal degree when the community size is larger than 40. 

The results are encouraging and showing that our methods can detect communities with similar internal degree and smaller community in geographic span.

\subsection{Gowalla Network}

The second real world network is Gowalla. From the analysis in section \ref{sec:network}, we know that compared with the Twitter network, the geographic information in Gowalla has greater influence on the network structure. So the Gowalla network is more suitable to use our community detection methods. 

From Figure \ref{fig:gowalladis}, we can see that our methods have a strong effect on limiting the geographic span of communities. Both the two methods can keep the span around or less than 200 kilometers. Especially for the connection locality method, even when the community size is very large, it can still keep the geographic span in a small range.

Another important observation is that in the highly geographically influenced networks, our method can also improve the network tightness in the communities. Figure \ref{fig:gowalladegree} shows the results of the average internal degree. The performances of these algorithm are similar to the case on the Twitter network. The different is that in the Twitter network, the connection locality method performs worse than the other two methods. But on the Gowalla network, it performs much better. This phenomenon illustrates that on the high geographically influenced networks, our method can improve the quality of the detection results on both geographic span and the tightness inside communities.

\section{Conclusion} \label{sec:conc}

In this paper, we studied the algorithms used in community detection. We argue that finding communities with small geographic span is important for many application domains. We analyzed two real datasets and found that they have different level of locality. We propose a new community detection method that keep the communities  in small range of areas while maintaining the connection closeness of the nodes in the communities. We performed extensive experiments on both synthetic and real world datasets. Results show that the proposed method find communities with nodes distributing in a smaller area compared with the traditional methods and having the similar or higher tightness on network connections. In our future work, we would like to explore low cost community detection algorithm utilizing the property of locality of nodes in communities. 


\bibliography{cite}
\bibliographystyle{plain}

\end{document}